\documentclass[conference]{IEEEtran}
\usepackage{dirtree}
\usepackage{graphicx}

\usepackage{cite}
\ifCLASSINFOpdf
  \usepackage[pdftex]{graphicx}
\else
\fi
\begin{document}
\title{NEARBY Platform for Detecting Asteroids in Astronomical Images Using Cloud-based Containerized Applications}

%\author{
%
%\IEEEauthorblockN{Victor Bacu, Adrian Sabou, Teodor Stefanut, Dorian Gorgan}
%\IEEEauthorblockA{Computer Science Department, \\
%Technical University of Cluj-Napoca \\ Cluj-Napoca, Romania \\
%victor.bacu, adrian.sabou, teodor.stefanut, dorian.gorgan@cs.utcluj.ro}
%\and
%\IEEEauthorblockN{Ovidiu Vaduvescu}
%\IEEEauthorblockA{Isaac Newton Group of Telescopes (ING),\\
%Santa Cruz de la Palma, Canary Islands, Spain\\
%Instituto de Astrofisica de Canarias (IAC),\\
%La Laguna, Canary Islands, Spain}
%ovidiu.vaduvescu@gmail.com}


\author{\IEEEauthorblockN{Victor Bacu\IEEEauthorrefmark{1},
Adrian Sabou\IEEEauthorrefmark{1},
Teodor Stefanut\IEEEauthorrefmark{1}, 
Dorian Gorgan\IEEEauthorrefmark{1} and
Ovidiu Vaduvescu\IEEEauthorrefmark{2}}
\IEEEauthorblockA{\IEEEauthorrefmark{1}Computer Science Department, Technical University of Cluj-Napoca\\Cluj-Napoca, Romania\\Email: \{victor.bacu, adrian.sabou, teodor.stefanut, dorian.gorgan\}@cs.utcluj.ro}
\IEEEauthorblockA{\IEEEauthorrefmark{2}Isaac Newton Group of Telescopes (ING), Santa Cruz de la Palma, Canary Islands, Spain\\
Instituto de Astrofisica de Canarias (IAC), La Laguna, Canary Islands, Spain\\
Email: ovidiu.vaduvescu@gmail.com}}


\maketitle

\begin{abstract}

The continuing monitoring and surveying of the nearby space to detect Near Earth Objects (NEOs) and Near Earth Asteroids (NEAs) are essential because of the threats that this kind of objects impose on the future of our planet. We need more computational resources and advanced algorithms to deal with the exponential growth of the digital cameras' performances and to be able to process (in near real-time) data coming from large surveys. This paper presents a software platform called NEARBY that supports automated detection of moving sources (asteroids) among stars from astronomical images. The detection procedure is based on the classic "blink" detection and, after that, the system supports visual analysis techniques to validate the moving sources, assisted by static and dynamical presentations.

\end{abstract}


% \fi
\IEEEpeerreviewmaketitle

\section{Introduction}

Near Earth Objects (NEOs) and and Near Earth Asteroids (NEAs) represent a vast research direction in the field of astrophysics and space sciences. Their discovery is important for understanding the nearby universe and also for the potential threat coming from their proximity to Earth. The discovery of NEOs and NEAs is a challenging task due to multiple factors, such as their small size, faint magnitudes and velocity \cite{Vaduvescu}.

Astronomical images are captured using mosaic cameras (containing an array of CCDs). This raw data should be transformed/modified before it can be used for asteroids detection. This process is known as data reduction and removes instrumental signatures, masks cosmic rays and makes photometric and astrometric calibration. 

An efficient computing infrastructure is needed to achieve very fast data reduction for rapid detection and validation. The solution presented in this paper targets the following requirements:
\begin{itemize}
\item Process and analyze multidimensional data in order to detect and identify moving objects in astronomical images;
\item Visual analysis of the processed images and human validation of the moving sources, assisted by static and dynamical presentations;
\item Flexible description of the processing pipeline;
\item Adaptive processing and detection over high-performance, cloud-based, computation infrastructures.
\end{itemize}

This paper is structured as follows: Section 2 presents some related works on the presented topic, while Section 3 presents the pipeline used to reduce astronomical images and to detect asteroids. The next sections describe in detail the design and implementation of the proposed solution. The last sections describe the validation of the NEARBY platform and conclude the ideas presented in this paper.

\section{Related Works}

Even though most major surveys have written their own automated system for asteroid detection, information on such existing applications is very difficult to come by, since virtually none were made public.

Research on automated asteroid detection systems can be traced back as early as 1992, when Scotti et al. \cite{1992acm..proc..541S} conducted a survey for NEAs with a TK2048 CCD in the scanning mode. The image data were transfered to an on-site computer that implemented a ``Moving Object Detection Program'' using 3 Sparc CPUs to look for streaked images of nearby asteroids and to identify sets of images that displayed consistent motion.

The Near-Earth Asteroid Tracking (NEAT) program was among the first to implement a fully automated system for detecting asteroids, including controlling the telescope, acquiring wide-field images, and detecting NEOs using on-site computing power provided by a Sun Sparc 20 and a Sun Enterprise 450 computer \cite{1538-3881-117-3-1616}.

Petit et al. \cite{doi:10.1111/j.1365-2966.2004.07217.x} developed a highly automated moving object detection software package. Their approach maintains high efficiency while producing low false-detection rates by using two independent detection algorithms and combining the results, reducing the operator time associated with searching huge data sets.

The first automatic detection improved algorithm, that cleans all stars in all individual images before combining them, was developed by Yanagisawa et al. \cite{doi:10.1093/pasj/57.2.399}. The algorithm uses many CCD images in order to detect very dark moving objects that are invisible on a single CCD image. It was applied for the first time on a very small 35-cm telescope to discover few asteroids up to mag 21 upon combination of 40 individual images exposed each for 3 minutes.

Pan-STARRS Moving Object Processing System (MOPS) \cite{10.1086/670337} is a modern software package that produces automatic asteroid discoveries and identifications from catalogs of transient detections from next-generation astronomical survey telescopes. MOPS achieves \textgreater 99:5\% efficiency in producing orbits from a synthetic but realistic population of asteroids whose measurements were simulated for a Pan-STARRS4-class telescope.

Besides major surveys, automated asteroid detection systems were also created by amateurs and small private surveys such as the TOTAS survey \cite{KOSCHNY2015305} carried out with the ESA-OGS 1m telescope, lead by ESA, or the work of Allekotte et al. \cite{10.1007/978-3-642-41827-3_15}.

Cop\^{a}ndean et al. \cite{8117033} propose an automated pipeline prototype for asteroids detection, written in Python under Linux, which calls some 3rd party astrophysics libraries. The current version of the proposed pipeline prototype is tightly coupled with the data obtained from the 2.5 meters diameter Isaac Newton Telescope (INT) located in La Palma, Canary Islands, Spain and represents the basis on which the NEARBY platform is built upon.

\section{Asteroids Detection Application (MOPS)}

An automated method to detect asteroids using just a single astronomical image is impossible. Most applications are trying to identify objects (called sources) from astronomical images (stars, asteroids, noise, etc.) and then try to link them to make a valid trajectory (considering a linear movement by constant speed). 

A survey is a collection of astronomic images taken by pointing the telescope to cover a large sky area. This area is composed by multiple fields. In order to maximize the potential to identify asteroids, a minimum of 4 repetitions per field is required. For small surveys (containing few fields observed in each night) and small mosaic cameras (just few CCDs) a manual detection procedure for discovering asteroids could be possible. However, for larger surveys, covering much more fields (sky area) and observation nights, an automated moving object pipeline is mandatory. NEARBY platform implements the Moving Object Processing System (MOPS), combining different functionalities provided by various third-party applications in order to extract and identify asteroids from a sequence of astronomical images. MOPS was originally mentioned in \cite{10.1086/670337} and used in the Pan-STARRS system.

The current pipeline integrated in our system \cite{8117033} is a classical blinking moving detection, that has the advantage of needing just a few images of the same field in order to produce good results. We need a minimum of 4 repetitions of the same field, but to increase the accuracy and the successful rate the number of repetitions should be higher, at least 5. The observed area in each field is constant in this situation. However, the fast moving objects (NEAs) will appear trailing (as longer lines, instead of star-like) during the exposure (typically 1 min), which makes them extremely difficult to detect automatically.

\begin{figure}[!t]
\centering
\includegraphics[scale=0.55]{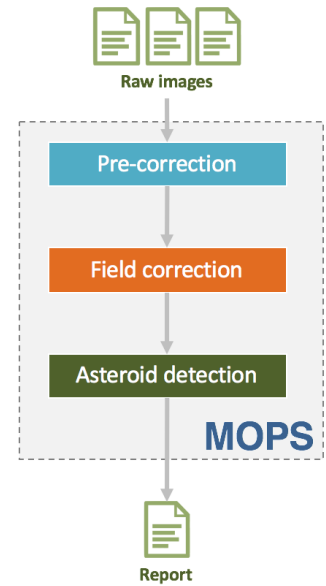} 
\caption{MOPS modules}
\label{fig:MOPS}
\end{figure}

Three modules are linked together in MOPS (see Figure \ref{fig:MOPS}), the first module receives as input the raw images and applies some pre-correction to them. The output is received by the next module dealing with field correction and extraction of potential objects (asteroids and stars). The last module starts from this list of objects and tries to group them to create valid asteroid trajectories considering a linear movement. The output of the last module is a report describing all the potential asteroids that were detected.

\textbf{Pre-correction module} takes care of correcting artifacts and bad pixels from the raw input. Flat files are used to correct some optical imperfections coming from the fact that usually telescopes distribute unevenly the light across the sensor and also contain dust. Bias files remove the CCD chip readout signal. The procedure consists in taking a large number of flat and bias images and combine them into a master flat/bias that will be applied to the raw images. This module is written in Python and uses IRAF \cite{IRAF} to generate a list of modified files that are sent to the next module. IRAF is a collection of software packages developed by the National Optical Astronomy Observatory (NOAO) aimed to reduce (process) astronomical images.

\textbf{Field correction module} deals with field distortions and realigns the raw images. Several steps are involved in this process:
\begin{itemize}
\item Modify raw data headers with the correct data (using IRAF \cite{IRAF})
\item Extract and build a catalog of astronomical objects (using SEXTRACTOR \cite{SEXTRACTOR})
\item Compute shifting function used to correct field distortion (using SCAMP \cite{SCAMP})
\item Resample images based on the shifting functions (using SWARP \cite{SWARP})
\item Extract and build a new catalog of astronomical objects after the images are corrected and resampled (using SEXTRACTOR)
\end{itemize}

\textbf{Asteroid detection module} has the main purpose of identifying asteroids trajectories from the elements detected by Sextractor in the processed images. It is important to underline the fact that, currently, the implementation of this module relies only on the features determined by Sextractor both on searching for valid asteroids trajectories and on reducing the number of false positives. No metadata information like position of stars and galaxies or list of known asteroids that should cross through the analyzed field at the observation time are used. All the detections rely solely on the captured images. The general idea of the algorithm implemented in this module is based on the fact that an asteroid, during the observation window, moves with constant speed and on a trajectory that can be considered linear. As a consequence, the algorithm tries to identify which of the sources identified by Sextrator can be combined with respect to the conditions above. The main challenge of this approach is generated by the fact that sources positioning has an error that is often comparable with the distance traveled by the asteroid between two consecutive captures. This enforces relaxations of the speed and trajectory angle computations which, in turn, favors the false positives.

\section{NEARBY Platform}

\subsection{Architecture}
\begin{figure}[!t]
\centering
\includegraphics[scale=0.5]{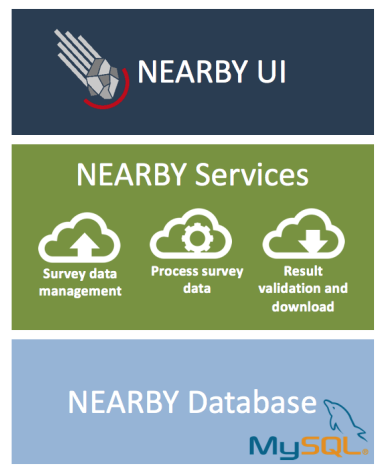} 
\caption{NEARBY System Architecture}
\label{fig:Architecture}
\end{figure}

The NEARBY platform architecture uses a three-tier model to provide flexible and reusable components, having presentation, logic and data storage tiers (see Figure \ref{fig:Architecture}). The infrastructure supporting this architecture is based on Kubernetes and Docker containers. It encapsulates every tier inside different containers, making it more easily to change and to adapt the configuration to a particular use case. The database level supports data persistence and uses a MySQL server. The logic level exposes all the functionality as REST services. These services are consumed by the presentation level, which is a web-based graphical user interface.

\subsubsection{Presentation level}
The NearbyUI allows users to interact with the service level in an intuitive and flexible way. It is built using HTML5 and Bootstrap and it can be accessed through web browsers. Most of the applications dealing with processing of astronomical data are desktop based applications. Our solution moves the presentation tier into the Cloud (not just the logic tier), allowing in this manner multiple users to simultaneous interact with the application, or making collaborative work on reducing and validating potential asteroids easier. Data management component allows users to upload surveys, raw images, to modify configuration files and to download results. Validating asteroids after the automatic discovery is performed through a dedicated user interface, allowing users to visualize an animated sequence of images of asteroid detection and to validate/invalidate this detection. 

\begin{figure*}[hbtp]
\centering
\includegraphics[scale=0.6]{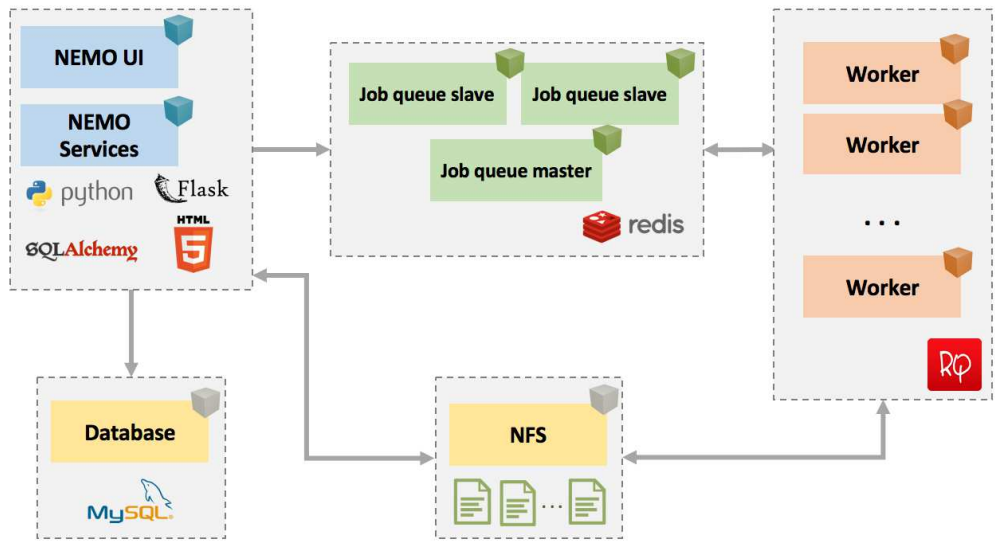} 
\caption{NEARBY Components}
\label{fig:Components}
\end{figure*}

\subsubsection{Service level}
NearbyServices exposes all the functionality as REST web services that are consumed by the top level (NearbyUI).  It is built using Flask \cite{FLASK}, a microframework for Python. It uses a modular approach, having defined services for surveys, nights, fields and projects (these notions are presented in the following section). Basically each service module supports the four basic functions of persistent storage, namely create, read, update and delete. The services level consists of three main modules, data module (survey data management), execution module (processing survey's data) and validation module (human validation of the identified asteroids).

\subsubsection{Data level}
NearbyDB stores all the information related to experiments, nights, fields and projects. After the automatic detection module identifies a set of potential asteroids this data is also stored in the database, allowing users to validate or invalidate it at a later time. Based on the list of validated asteroids the system will generate a report that will be sent to a validation authority. 

\subsection{Data storage}

A survey contains a collection of astronomical images gathered with the same telescope during a continuous time interval. This data has a hierarchical structure with the survey as the root level. Under each survey the data is grouped into nights. Each night contains the collection of observations done during the night. During a night the telescope pointing is changing in order to maximize the observed sky area. Each such telescope pointing represents a field and the actual data is available at this level. A project represents an executable instance of a particular field containing all the input data (raw images) together with the user defined set of configurable parameters. The output of a project is represented by a collection of identified asteroids which should be later validated or invalidated by the user.

The actual data (raw images) is stored in a directory structure that maps on the hierarchical structure of a survey. In a database we store the metadata related to this data, allowing us to easily retrieve various information about a particular survey.

The database is stored in a MySQL server instance and the link to it is performed through SQLAlchemy library, which is an Object Relational Mapper providing an efficient and high-performing database access. Each database table is mapped to a Python class describing that particular table. In the following paragraphs we will briefly present each of the database tables in which the data is stored. 

\textit{Survey} groups all the observations done with the same telescope. The user can specify the experiment name and a short description. A unique name is generated, representing the directory name in which all data will be stored. This info is not visible to the user and allows us to change the internal directory structure if needed. Each survey could have one or more observation nights.

\textit{Night} groups all the observations based on the acquisition date. As in the case of experiment the user can specify the night name and a short description. There is a back reference from night to experiment to simplify data access. Under each night we have one or more fields. Multiple nights can be defined for one experiment.

\textit{Field} represents a particular area that is observed during one night. The user can specify the field name and a short description. We have a back reference to the night to which this field belongs. Multiple fields can be defined for one night.

\textit{Project} can be thought of as an executable field. Here the user specifies a set of parameters in order to be able to process the raw data and to obtain a set of potential asteroids from this field. The user can specify a project name and a short description. The status field will be updated accordingly to the state in which the execution is. There is a back reference to the field on which the current project was created on. Multiple projects can be defined for one field.

\textit{AstObject} represents one potential asteroid identified in one image. The database contains the description of the potential asteroid by MJD (Modified Julian Date), RA (Right ascension), Dec (Declination) and Magnitude. Each such detection has a small image attached, centered on the asteroid and is used to display an animation of the asteroid’s movement.

\textit{Report} groups a set of potential asteroids identified in one field. This report is linked to the project and also to the author of it. It is possible to link multiple reports. That means the users can collaboratively work on the previous reports in order to enhance it. The link between Report and AstObject is done by Report\_Object. It contains reference to the Report and to the AstObject and also the status of the potential asteroid. This status is update by users and can be selected between VALID and INVALID.

As stated before, the actual data is stored in a directory structure that maps on the hierarchical structure of a survey. This hierarchical structure organize and group data on surveys, nights and fields in order to reduce data redundancy.

\dirtree{%
.1 Survey.
.2 Input.
.3 BadPix.
.2 Night.
.3 Input.
.4 Bias.
.4 Flat.
.3 Field.
.4 Input.
.5 Row.
.4 Project.
.5 Params.
.5 Build.
}

At survey level we store the map (BadPix) representing the camera's pixels giving incorrect response and one or more directories representing observations done in different nights. The fields gathered during one night share the same set of images (Bias and Flat) stored under the current night. At field level we store the raw data (observation images). The project directory contains the parameters used to run a MOPS instance on the particular set of input data (Params) and the results of the execution (Build).

\subsection{Execution}

Execution module is responsible for running MOPS on astronomical images from the data repository and for this is using various third-party applications dealing with image corrections, extraction of potential asteroids, resampling and realigning images, etc. The execution module was design to address the followings:
\begin{itemize}
\item scalability –- Kubernetes supports expansion or shrinking of the pods that are available in a deployment without affecting the ongoing processing. In this manner, at surveys time we can process in real time all the field by increasing the number of computational nodes that perform the detections. On the other hand, we can limit the number of the computational nodes available by turning off some of them and use these resources for other applications;
\item fault tolerance –- in large computational clusters the risk of having a hardware failure is high and in this situations the system should adapt. The pods are automatically restarted if something goes wrong. Obviously in this situation the execution should be restarted based on the fact that the data could be inconsistent;
\item efficiency -– scheduling policies could be imposed based on some metrics that the system should relate to. The pods are scheduled on the most suitable nodes from the Kubernetes cluster. If the data communication cost is too high, it could be a solution to move the processing closer to the data. This could be become a valid solution if the cluster grows to be more geographically distributed.
\end{itemize}

The execution module generates job description and sends them to a job queue. Worker nodes listening to this job queue will process the data. As soon as the job is stored on the queue if a worker node is available it will fetch the job info and it will start the execution. If no worker nodes are available, then the jobs will stay in queue but only for a limited amount of time. We chose this approach in order to limit the data that is stored in the job queue. 

After having uploaded all the necessary files through UI the user can start the processing in order to detect asteroids. For this he has to create a new project. This project contains all the configuration files needed by different applications that are used in asteroid detection. When a new project is created a set of predefined configuration files are copied to the project location. The user has the possibility to use these files or to modify them. The service level will generate a job description based on the project selected by the user and will send this description to a job queue. Different worker nodes will fetch jobs from there, execute them and in the end will store the results that will be available to the user. The job execution is monitored and the job details that are stored in the database  will be updated.

We implemented the procedure of queueing jobs and processing them on different worker nodes using RQ (Redis Queue). The queue is backed by Redis, an open-source in-memory database. A job is actually a Python object and is represented by a function that will be invoked by a worker node. All the components from Figure \ref{fig:Components} run inside a different Kubernetes deployment. Worker nodes listen on different queues and fetches jobs from there. The survey’s data is shared by the service and worker deployments through a distributed file system (NFS). 

\begin{table}[]
\centering
\caption{Evaluation results}
\label{table:Results}
\begin{tabular}{|c|c|c|c|}
\hline
Field                      & Total detections & Valid detections & Execution time \\ \hline
E101                       & 33              & 18               & 00:07:45       \\ \hline
\multicolumn{1}{|l|}{E102} & 47              & 24               & 00:08:02       \\ \hline
\multicolumn{1}{|l|}{E103} & 46              & 19               & 00:07:43       \\ \hline
E104                       & 27              & 10               & 00:08:20       \\ \hline
E105                       & 72              & 21               & 00:08:18       \\ \hline
E106                       & 63              & 13               & 00:07:52       \\ \hline
E107                       & 120             & 26               & 00:08:44       \\ \hline
E108                       & 43              & 15               & 00:08:15       \\ \hline
E109                       & 48              & 21               & 00:08:02       \\ \hline
E110                       & 35              & 17               & 00:08:08       \\ \hline
E111                       & 54              & 13               & 00:08:30       \\ \hline
E112                       & 40              & 19               & 00:08:05       \\ \hline
\end{tabular}
\end{table}

\begin{figure*}[hbtp]
\centering
\includegraphics[scale=0.8]{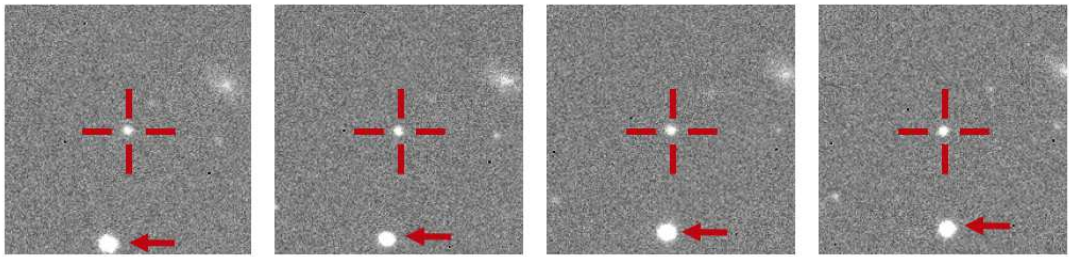} 
\caption{Asteroid detection}
\label{fig:Detections}
\end{figure*}

\section{Testing}

This section covers the evaluation of the proposed software platform. The testing environment has the following resources: 3 IBM Blade servers, powered by Intel Xeon Processor E5-2630 v2 6C and 64 GB memory, and 1 IBM Blade server, powered by Intel Xeon Processor E5-2630 v2 6C and 16 GB memory. The Cloud infrastructure is built using OpenStack. On top of this infrastructure we build a Kubernetes cluster that manages the containerized applications.

The INT telescope and WFC mosaic camera used in the evaluation of our solution covers 0.27 sq.deg for each pointing with 4 CCDs. The raw images taken during the survey are in the FITS format. For the INT telescope the input data consists of 4 sets of raw mosaic images in standard astronomic multi-extension FITS format (each CCD image is inside a FITS plane).

The testing case is a survey of 3 observation nights, containg 12, 20 and 24 fields. To evaluate the system, we measured the performance and also the number of valid detections. A 5 worker nodes pool was used to run MOPS. The average execution time varies between 7 and 14 minutes. Table \ref{table:Results} presents the results for the first night. The execution time is influenced by the NFS server through which all the files are accessed and which could become a bottleneck for the application. During the execution the fields were copied to the NEARBY platform, this also influencing the total execution time. The first set of data contained 12 fields, reduced automatically by NEARBY platform in around 24 minutes. A local execution would take around 100 minutes. The obtained speed-up compared with the local execution makes this solution viable for reducing a large set of data. 

From the valid detection point of view, the results are also promising. Table \ref{table:Results} presents the number of potential asteroids detected by MOPS (total detections column) and the number of asteroids that were correctly identified by MOPS (valid detections). To validate results we need reports containing asteroids detected by humans on the same field. It is worthwhile to mention that not all of the asteroids detected by humans can actually be detected automatically. The reason for this is that the asteroid detection module works with objects (sources) identified by a third-party application (Sextractor) and in some cases these asteroids are not detected (due to bad pixels, faint objects, etc.). Figure \ref{fig:Detections} shows what a detection looks like. The image is centered on asteroid and the stars will move on the animated sequence of images. It can be seen that the arrow showing a star is actually moving. Compared with human detection MOPS identifies at least 70\% but this value could be increased by better tweaking the parameters influencing the automatic objects detection. The number of false detections on this survey is still high but this result is influenced by some factors such as the camera used in this survey which contains a lot of bad pixels.

\section{Conclusions}

The discovery of NEOs and NEAs is a challenging task due to multiple factors, such as their small size, faint magnitudes and velocity. The solution presented in this paper allows users to process and analyze multidimensional data in order to detect and identify moving objects in astronomical images. Visual analysis of the processed images and presentation of the moving sources assisted by static and dynamical animations supports human validation. The evaluation shows good performance both for the number of valid detected asteroids and for the computational gain that this solution provides.


\section*{Acknowledgment}

This research is supported by ROSA (Romanian Space Agency) by the Contract CDI-STAR 192/2017, NEARBY - Visual Analysis of Multidimensional Astrophysics Data for Moving Objects Detection. The data used to test the NEARBY platform was available by observations made with the Isaac Newton Telescope (INT) operated on the island of La Palma by the Isaac Newton Group (ING) in the Spanish Observatorio del Roque de los Muchachos (ORM) of the Instituto de Astrofisica de Canarias (IAC).



\bibliographystyle{IEEEtran}
\bibliography{IEEEabrv,paper}
%\begin{thebibliography}{1}


%\end{thebibliography}

\end{document}